\documentclass[12pt,preprintnumbers]{revtex4}
\usepackage{latexsym,graphicx}

\newcommand{\be}{\begin{equation}}
\newcommand{\ee}{\end{equation}}
\newcommand{\mn}{{\mu\nu}}
\newcommand{\m}{{\rm m}}
\newcommand{\gs}{{G_*}}
\newcommand{\gc}{{G_{\mathrm c}}}
\newcommand{\gn}{{G_{\mathrm N}}}
\newcommand{\adota}{H}
\newcommand{\adotdota}{\left(\frac{\ddot{a}}{a}\right)}

\begin{document}
\preprint{EFI-04-26}
\preprint{hep-th/0407149}

\title{Lorentz-Violating Vector Fields Slow the Universe Down}
\author{Sean M. Carroll$^{1,3}$ 
and Eugene A. Lim$^{2,3}$}

\affiliation{$^{1}$Department of Physics,\\
$^{2}$Department of Astronomy and Astrophysics,\\
$^{3}$Enrico Fermi Institute, and
Kavli Institute for Cosmological Physics,\\
University of Chicago, 5640 S. Ellis Avenue, Chicago, IL 60637 \\
{\tt carroll@theory.uchicago.edu, elim@oddjob.uchicago.edu}}

\begin{abstract}
We consider the gravitational effects of a single, fixed-norm, Lorentz-violating timelike vector field.  In a cosmological background, such a vector field acts to rescale the effective value of Newton's constant.  The energy density of this vector field precisely tracks the energy density of the rest of the universe, but with the opposite sign, so that the universe experiences a slower rate of expansion for a given matter content. This vector field similarly rescales Newton's constant in the Newtonian limit, although by a different factor. We put constraints on the parameters of the theory using the predictions of primordial nucleosynthesis, demonstrating that the norm of the vector field should be less than the Planck scale by an order of magnitude or more.
\end{abstract}

\maketitle
\newpage

\section{Introduction}

Lorentz invariance is a fundamental requirement of the standard model of particle physics, verified to high precision by many tests \cite{Kostelecky:2001xz}. Nevertheless, there exist good reasons to push tests of this symmetry to increasing levels of precision. For example, string theory predicts that we may live in a universe with non-commutative coordinates \cite{Connes:1998cr}, leading to a violation of Lorentz invariance \cite{Carroll:2001ws}. Furthermore, astrophysical observations suggests the presence of high energy cosmic rays above the Greisen-Zatsepin-Kuzmin cutoff \cite{GZK}, results which may be explained by a breaking down of Lorentz invariance \cite{Chisholm:2003bu,Dubovsky:2001hj,Coleman:1998ti,Vankov:2002gt,Kifune:1999ex,Zee:1981sy,Aloisio:2000cm,Bertolami:1999dc,Bertolami:1999da,Alfaro:2002ya,Coleman:1997xq,Gagnon:2004xh}. More generally, our ability to test Lorentz invariance to very high precision provides a unique window into unknown effects at the Planck scale \cite{Boggs:2003kx}.

A straightforward method of implementing local Lorentz violation in a gravitational setting is to imagine the existence of a tensor field with a non-vanishing expectation value, and then couple this tensor to gravity or matter fields.  The simplest example of this approach is to consider a single timelike vector field with fixed norm. A special case of this theory was first introduced as a mechanism for Lorentz-violation by Kostelecky and Samuel in \cite{Kostelecky:1989jw}. In this paper, we will consider the more general theory suggested by Jacobson and Mattingly in \cite{Jacobson:2001yj,Kostelecky:2003fs}.  (In a different context, Bekenstein has proposed a theory of gravity with a fixed-norm vector in order to mimic the effects of dark matter \cite{Bekenstein:2004ne}; also, studies of vector fields in a cosmological setting without the fixed norm have been done elsewhere \cite{Will:ns,Armendariz-Picon:2004pm,Ford:1989me, Dolgov:1982qq, Kiselev:2004py, Kiselev:2004vy, Kiselev:2004bh}.) 

This vector field picks out a preferred frame at each point in spacetime, and any matter fields coupled to it will experience a violation of local Lorentz invariance \cite{Colladay:1998fq,Carroll:1990vb}.  For the purposes of laboratory tests, it suffices to take the vector field to be a fixed element of a background flat spacetime, with constant components in inertial coordinates.  In curved spacetime, however, there is no natural generalization of the notion of a constant vector field (since $\nabla_\mu u^\nu=0$ generically has no solutions); we must therefore allow the vector to have dynamics, and fix its norm by choosing an appropriate action for the field.

In this paper we investigate the gravitational effects of such vector fields, especially in the context of cosmology.  We find a non-trivial impact on the evolution of the universe, namely to decrease the effective value of Newton's constant relative to that measured in the Solar System, resulting in a slowing down of the expansion rate for any fixed matter content.  

We begin by deriving the basic equations of motion for the most general theory of a fixed-norm vector field $u^\mu(x)$ with an action that is at most second order in the vector and its derivatives.  We study the resulting theory in the Newtonian limit, showing that it acts to rescale the gravitational constant.  We then study the Friedmann equations in this theory, again finding a rescaling of the gravitational constant but by a different factor.  This fact allows us to use the precise predictions of Big Bang nucleosynthesis to put a limit on the parameters of this theory.  We are able to rule out the possibility that the norm of the vector is as large as the Planck scale.

In the course of this paper we refer to some results from a companion paper that focuses on the cosmological behavior of perturbations in the vector field \cite{Lim:2004js}.

\section{The Equations of Motion}

The theory we consider consists of a vector field $u^\mu$ minimally coupled to gravity, with an action of the form
\begin{equation}
S=\int d^4x \sqrt{-g} \left(\frac{1}{16\pi \gs }R + {\cal{L}}_u +{\cal{L}}_m \right) . \label{theoryaction}
\end{equation}
The parameter $\gs$ is given an asterisk subscript to emphasize that it will not be equal to the value of Newton's constant that we would measure either in the Solar System or in cosmology. ${\cal{L}}_u$ is the vector field Lagrange density while ${\cal{L}}_m$ denotes the Lagrange density for all other matter fields. The Lagrange density for the vector consists of terms quadratic in the field and its derivatives:
\begin{equation}
  {\cal{L}}_u = -\beta_1 \nabla^\mu u^\sigma \nabla_\mu u_\sigma - \beta_2
  (\nabla_\mu u^\mu)^2 - \beta_3 \nabla^\mu u^\sigma \nabla_\sigma u_\mu
  + \lambda(u^{\mu} u_{\mu} +m^2)\ .
\end{equation}
Here the $\beta_i$'s are dimensionless parameters of our theory, and $\lambda$ is a Lagrange multiplier field; the vector field itself has a dimension of mass. This is a slight simplification of the theory introduced by Jacobson and Mattingly \cite{Jacobson:2001yj} (with a slightly different parameterization), as we have neglected a quartic self-interaction term $(u^{\rho}\nabla_{\rho}u^{\mu})(u^{\sigma}\nabla_{\sigma}u_{\mu})$.  It will be convenient in what follows to express the Lagrangian as
\begin{equation}
{\cal{L}}_u = K^{\mu\nu}{}_{\sigma\rho}\nabla_{\mu} u^{\sigma} \nabla_{\nu} u^{\rho}+ \lambda(u^{\mu} u_{\mu} +m^2)\ ,
\end{equation}
where
\begin{equation}
K^{\mu\nu}{}_{\sigma\rho}=-\beta_1 g^{\mu\nu}g_{\sigma\rho}-\beta_2\delta^{\mu}_{\sigma} \delta^{\nu}_{\rho}-\beta_3\delta^{\mu}_{\rho}\delta^{\nu}_{\sigma}.
\end{equation}

We also define a current tensor $J^{\mu}{}_{\sigma}$ via
\begin{equation}
J^{\mu}{}_{\sigma}\equiv K^{\mu\nu}{}_{\sigma\rho}\nabla_{\nu}u^{\rho}.
\end{equation}
Note that the symmetry of $K^{\mu\nu}{}_{\sigma\rho}$ means that $J^{\nu}{}_{\rho}=K^{\mu\nu}{}_{\sigma\rho}\nabla_{\mu}u^{\sigma}$. With these definitions, the equation of motion obtained by varying the action with respect to $u^\mu$ is
\begin{equation}
\nabla_{\mu}J^{\mu\nu}=\lambda u^{\nu}. \label{EOMforu}
\end{equation}
The equation of motion for $\lambda$ enforces the fixed norm constraint
\begin{equation}
  u_{\mu}u^{\mu}=-m^2. \label{FixedNorm}
\end{equation}
Choosing $m^2>0$ ensures that the vector will be timelike.

The value of $\lambda$ is determined by the equation of motion for the vector field.  This can be seen by decomposing the vector equation (\ref{EOMforu}) into components along and perpendicular to $u^\mu$.  Multiplying both sides of (\ref{EOMforu}) by $u_\nu$ and using (\ref{FixedNorm}), we find
\be
  \lambda = -{1\over m^2}u_\nu \nabla_\mu J^\mn .
  \label{lambda}
\ee
We can also project into a subspace orthogonal to $u^\mu$ by acting the projection tensor $P^\rho{}_\nu = m^{-2}u^\rho u_\nu + \delta^\rho_\nu$ on  (\ref{EOMforu}) to obtain
\be
  u^\rho u_\nu \nabla_\mu J^\mn + m^2 \nabla_\mu J^{\mu\rho}=0 .
  \label{projectedu}
\ee
This equation determines the dynamics of $u^\mu$, subject to the fixed-norm constraint.

The stress-energy tensor for the vector field is
\be
T_{\mu\nu}= -{2\over \sqrt{-g}}{\delta S_u \over \delta g^\mn} =
-2\frac{\partial {\cal{L}}_u}{\partial g^{\mu\nu}}+g_{\mu\nu}{\cal{L}}_u .
\ee
In taking the variation, it is important to distinguish which variables are independent.  Our dynamical degrees of freedom are the inverse metric $g^\mn$ and the contravariant vector field $u^\mu$.  If we denote by $\delta_g$ the change in a quantity under the variation $g^\mn \rightarrow g^\mn + \delta g^\mn$, the variation of the covariant vector $u_\sigma$ is given by
\be
  \delta_g u_\sigma = \delta_g (g_{\rho\sigma}u^\rho) = u^\rho
  \delta_g g_{\rho\sigma} = -u^\rho g_{\mu\rho}g_{\nu\sigma} \delta g^\mn
  = -u_\mu g_{\nu\sigma}\delta g^\mn ,
\ee
where we have used the identity
\be
  \delta_g g_{\rho\sigma}=-g_{\mu\rho}g_{\nu\sigma} \delta g^\mn .
\ee
In contrast, the variation of the contravariant vector $u^\rho$ with respect to the metric simply vanishes,
\be
  \delta_g u^\rho = 0 .
\ee

The variation of the Lagrange density ${\cal{L}}_u$ with respect to the metric, keeping $u^{\mu}$ fixed, is 
\begin{eqnarray}
\delta_g {\cal{L}}_u= K^{\mu\nu}{}_{\sigma\rho}\delta_g ( \nabla_{\mu} u^{\sigma})\nabla_{\nu} u^{\rho} + K^{\mu\nu}{}_{\sigma\rho}\nabla_{\mu} u^{\sigma} \delta_g(\nabla_{\nu} u^{\rho}) \nonumber\\
+\delta_g (K^{\mu\nu}{}_{\sigma\rho})\nabla_{\mu} u^{\sigma} \nabla_{\nu} u^{\rho}+\lambda \delta_g (g_{\mu\nu} u^{\mu} u^{\nu}). \label{VariationofL}
\end{eqnarray}
The first and second terms on the right hand side are identical due to the symmetry properties of the $K^{\mu\nu}{}_{\sigma\rho}$ tensor. To compute them we note that 
\begin{equation}
\delta_g (\nabla_{\mu} u^{\sigma})= \delta_g(\partial_\mu u^\sigma + \Gamma^\sigma_{\mu\rho}u^\rho) = (\delta_g \Gamma^{\sigma}_{\mu\rho}) u^{\rho} .
\end{equation}
The variation of the Christoffel symbol yields
\be
\delta_g \Gamma^{\mu}_{\nu\sigma} = -\frac{1}{2}\left(g_{\sigma\rho}\nabla_{\nu} \delta g^{\mu\rho}+g_{\nu\rho}\nabla_{\sigma} \delta g^{\mu\rho}+g^{\mu\rho}\nabla_{\rho} \delta_g g_{\nu\sigma}\right),
\ee
from which we obtain (ignoring total derivatives)
\begin{equation}
K^{\mu\nu}{}_{\sigma\rho}  \nabla_{\nu} u^{\rho}\delta_g (\nabla_{\mu} u^{\sigma})=\frac{1}{2}\left[\nabla_{\mu}(u_{(\alpha} J^{\mu}{}_{\beta)})+\nabla_{\mu} (u^{\mu}J_{(\alpha\beta)})-\nabla_{\mu}(u_{(\alpha} J_{\beta)}{}^{\mu})\right]\delta g^{\alpha\beta} .
\end{equation}
The third term in the right hand side of the variation Eqn. (\ref{VariationofL}) is simply
\begin{equation}
\delta_g K^{\mu\nu}{}_{\sigma\rho} =-\beta_1(g_{\sigma\rho}\delta^{\mu}_{\alpha} \delta^{\nu}_{\beta}-g^{\mu\nu}g_{\alpha\sigma}g_{\beta\rho})\delta g^{\alpha\beta} ,
\end{equation}
and for the last term we get 
\begin{equation}
\delta_g (g_{\mu\nu} u^{\mu}u^{\nu})=- u_{\alpha}u_{\beta}\delta g^{\alpha\beta}.
\end{equation}

Putting everything together, the stress-energy tensor is
\begin{eqnarray}
T_{\mu\nu}^{(u)}&=&2\beta_1(\nabla_{\mu} u^{\rho}\nabla_{\nu} u_{\rho}-\nabla^{\rho} u_{\mu} \nabla_{\rho} u_{\nu})  \nonumber \\
&&-2[\nabla_{\rho}(u_{(\mu}J^{\rho}{}_{\nu)})+\nabla_{\rho}(u^{\rho} J_{(\mu\nu)}))-\nabla_{\rho}(u_{(\mu}J_{\nu)}{}^{\rho})] \nonumber \\
&&-2m^{-2}u_{\sigma}\nabla_{\rho} J^{\rho\sigma}u_{\mu} u_{\nu}+g_{\mu\nu}{\cal{L}}_u , \label{stressenergy}
\end{eqnarray}
where we have used Eqn. (\ref{lambda}) to replace $\lambda$. This was first derived in \cite{Jacobson:2004ts}.

In Minkowski space, a simple solution to the theory is a constant-field
configuration,
\be
  u^\mu = (m,0,0,0) .
\ee
It is easy to check that the stress-energy tensor for such a configuration vanishes identically.

\section{The Newtonian Limit}

We are ultimately interested in the potentially observable effects of the vector field on the expansion of the universe, which we will find to be a rescaling of the effective value of Newton's constant.  Such an effect would not truly be observable, however, if the same rescaling affected our measurements in the Solar System.  We therefore begin with an examination of the Newtonian limit, in which fields and sources are taken to be both static (no time derivatives) and weak (so that we may neglect terms beyond linear order).  This limit suffices to describe any laboratory measurements of the effective value of Newton's constant, which we will denote $\gn$. We note that our results are consistent with \cite{Eling:2003rd} where a more detailed analysis for a static spherically symmetric system without a source was done using the Eddington-Robertson-Schiff Post-Newtonian parameters.

The metric in the Newtonian limit takes the form
\begin{equation}
  ds^2=-(1+2\Phi({\vec x}))dt^2+(1-2\Psi({\vec x}))(dx^2 + dy^2 + dz^2).
  \label{weak}
\end{equation}
We will look for solutions in which the dynamical vector field $u^\mu$ is parallel to the timelike Killing vector $\xi^\mu = (1,0,0,0)$, so the only nonvanishing component will be $u^0$.  Given the fixed-norm condition, at linear order we have
\begin{equation}
  u^{0}=(1-\Phi)m .
\end{equation} 
The equation of motion (\ref{lambda}) gives us the solution for $\lambda$ at linear order,
\begin{equation}
\lambda(\vec{x})={\beta_3\nabla^2 \Phi},
\end{equation}
where $\nabla^2$ is the ordinary three-dimensional Laplacian.  It is then straightforward to verify that $u^\mu$ satisfies (\ref{projectedu}).  There may, of course, be solutions with nonzero spatial components of $u^{\mu}$, which we do not consider here. Spatial components were considered in \cite{Eling:2003rd}, which found that in a static spherically symmetric system, the vector field must have radial components for it to be consistent when non-linear effects are included.

A straightforward computation shows that the only nonvanishing component of the stress-energy tensor for the vector field (to linear order) is the effective energy density,
\begin{equation}
T_{00}^{(u)}= -2\delta\nabla^2\Phi ,
\end{equation}
where we have defined a parameter
\be
  \delta\equiv -\beta_1 m^2 
  \label{beta}
\ee
that will be convenient for later comparison with the cosmological results.

In addition to the vector field, we model the matter stress-energy tensor by a static distribution of dust; at lowest order, the only nonvanishing component is again the energy density,
\begin{equation}
T_{00}^{(\m)}= \rho_\m({\vec x}).
\end{equation}
This might describe the Earth, the Sun, or laboratory sources.  Note that we have assumed no coupling of the matter fields to the vector field. This means that we will not see any Lorentz violation in the dynamics of the matter fields, since they do not feel the presence of the preferred frame. We imagine that direct interactions can be safely included as a perturbation.

Einstein's equation in the presence of both matter and the vector field is
\be
  R_\mn - {1\over 2}Rg_\mn = 8\pi \gs (T_\mn^{(u)}+T_\mn^{(\m)}) ,
\ee
where $\gs$ is the parameter in the action (\ref{theoryaction}).  The spatial components of Einstein's equation for the metric (\ref{weak}) are
\begin{equation}
(\delta_{ij}\nabla^2 - \partial_i\partial_j)(\Phi-\Psi) = 0.
\end{equation}
Assuming boundary conditions such that both $\Phi$ and $\Psi$ vanish at spatial infinity, the unique solution is then
\be
  \Psi = \Phi .
\ee
The $00$ component yields
\be
  2\nabla^2\Phi = 8\pi\gs (T_{00}^{(u)}+T_{00}^{(\m)})
  = 8\pi\gs (-2\delta\nabla^2\Phi + \rho_\m).
\ee
This can be rewritten in the form of the Poisson equation,
\be
  \nabla^2 \Phi = 4\pi \gn \rho_\m ,
\ee
where we have defined
\be
  \gn = {\gs \over 1+8\pi \gs\delta}.
  \label{gndef}
\ee
This is the effective value of Newton's constant that we would actually measure via experiments within the Solar System. 

We see that local experiments in the Newtonian limit cannot themselves provide any constraints on the values of the parameters characterizing our vector-field theory.  Instead, there is a unique rescaling of the strength of the gravitational force; since $\gs$ is not directly measurable, however, we have no constraint on $\delta$.  To actually obtain a constraint we need to move beyond the Newtonian limit and consider cosmology.

\section{Cosmological Solutions}

Homogenous and spatially isotropic universes are described by the Robertson-Walker metric
\begin{equation}
ds^2=-dt^2+a^2(t)\left(\frac{1}{1-Kr^2}dr^2+r^2d\Omega^2\right)
\end{equation}
where the curvature parameter $K$ vanishes for a spatially flat universe. For such a metric to solve Einstein's equation in the presence of a fixed-norm vector field, the vector must respect spatial isotropy, at least in the background (though perturbations will generically break the symmetry). Thus the only component that the vector can possess is the timelike component. Using the fixed norm constraint (\ref{FixedNorm}), the components of the vector field are simply 
\begin{equation}
u^\mu=(m,0,0,0),
\label{cosmou}
\end{equation}
just as in flat spacetime.
{}From (\ref{lambda}) we then find that the Lagrange multiplier field takes the form
\begin{equation}
\lambda(t)=-3(\beta_1+\beta_2+\beta_3)\adota^2+3\beta_2\adotdota.
\end{equation}

We model the stress-energy tensor for the matter as a perfect fluid with energy density $\rho_\m$ and pressure $p_\m$,
\be
  T^{(\m)}_\mn = (\rho_\m + p_\m)n_\mu n_\nu + p_\m g_\mn ,
\ee
where $n_\mu$ is a unit timelike vector field representing the fluid four-velocity.  Although we refer to ``matter,'' this fluid may consist of any isotropic source; in particular, it can be a combination of different components with distinct equations of state.  The stress-energy for the vector field also takes the form of a perfect fluid,
with an energy density given by
\be
  \rho_u=-3\alpha H^2
  \label{rhou}
\ee
and a pressure
\be
  p_u=\alpha\left[\adota^2+2\adotdota\right] ,
\ee
where we have introduced the parameter 
\be
\alpha\equiv(\beta_1+3\beta_2+\beta_3)m^2 
  \label{alpha}
\ee
and $H=\dot{a}/a$ is the Hubble parameter.  It is easy to check that the vector field obeys the cosmological energy conservation relation $\dot{\rho_u}+3H(\rho_u+p_u)=0$. 

For the energy density (\ref{rhou}) of the background vector-field configuration to be positive requires $\alpha<0$. However, as is shown in a companion paper \cite{Lim:2004js}, insisting that the \emph{perturbed} degrees of freedom have a consistent quantum field theory as well as non-tachyonic and non-superluminal behaviour requires $(\beta_1+3\beta_2+\beta_3)\geq0$, implying the opposite condition:
\be
  \alpha \geq 0 .
\ee
Specifically, the following conditions hold for a well-defined theory:
\begin{eqnarray}
\beta_1 &>&0 \cr
(\beta_1+\beta_2+\beta_3)/\beta_1&\geq& 0 \cr
(\beta_1+\beta_2+\beta_3)/\beta_1&\leq& 1 \cr
\beta_1+\beta_3&\leq& 0 .
\label{conditions}
\end{eqnarray}
The first of these arises from the need for a positive-definite Hamiltonian for the perturbations; the next two from demanding non-tachyonic and subluminal propagation of the spin-0 degrees of freedom, respectively; and the last from insisting that gravity waves propagate subluminally. Together these conditions imply that 
\begin{eqnarray}
\beta_1+\beta_2+\beta_3&\geq&0 \\
\beta_2&\geq&0  
\end{eqnarray}
giving us our $\alpha\geq0$ constraint. In particular, the parameter choice $\beta_1=-\beta_3$ and $\beta_2=0$ corresponds to the vector field Lagrangian possessing only the anti-symmetric term 
\begin{equation}
{\cal{L}}_u=-\beta_1\nabla_{[\mu}u_{\nu]}\nabla^{[\mu}u^{\nu]}+\lambda(u^2+m^2)
\end{equation}                                                            
which was first considered in \cite{Colladay:1998fq}.                                                                                      
%In particular, the parameter choice $\beta_1=-\beta_3$ and $\beta_2=0$ corresponds to the vector field Lagrangian possessing only the anti-symmetric term 
%\begin{equation}
%{\cal{L}}_u=-\beta_1\nabla_{[\mu}u_{\nu]}\nabla^{[\mu}u^{\nu]}+\lambda(u^2+m^2).
%\end{equation} 

The background energy density of the vector field is therefore non-positive.  This is by no means a disaster, since the vector field itself in the background has no dynamical degrees of freedom, much like a negative cosmological constant. As we have mentioned earlier, the actual degrees of freedom corresponding to perturbations have a positive energy density and a positive norm on their Hilbert space (see also \cite{Arkani-Hamed:2003uy,Arkani-Hamed:2003uz,Arkani-Hamed:2004ar}). 

Including the ordinary matter fields, the Einstein equation $G_{\mu\nu}=8\pi \gs (T^{(u)}_{\mu\nu}+T^{(\m)}_{\mu\nu})$ becomes two equations,
\begin{eqnarray}
-\frac{3}{8\pi \gs }\left[H^2 +\frac{K}{a^2}\right]&=&3\alpha\adota^2 -\rho_\m  \label{Friedmann1}\\
-\frac{1}{8\pi \gs }\left[\adota^2+2\adotdota+\frac{K}{a^2}\right] &=& \alpha\left[\adota^2+2\adotdota\right]+p_\m . \label{Friedmann2}
\end{eqnarray}
Notice that, similar to the Newtonian limit case, the energy density $\rho_u$ is proportional to the square of the Hubble parameter. These Friedmann equations can be rewritten as
\begin{eqnarray}
H^2&=& \frac{8\pi\gc}{3}\rho_\m + \frac{\gc}{\gs }\frac{K}{a^2} \\
\frac{\ddot{a}}{a}&=&-\frac{4\pi \gc}{3}(\rho_\m+3p_\m) ,
\end{eqnarray}
where the effective Newton's constant is
\be
\gc\equiv\frac{\gs}{1+8\pi \gs\alpha} .
\label{gcdef}
\ee
Just as in (\ref{gndef}), the effect of the vector field is simply to change the observable value of Newton's constant, but now by a different parameter.  Since $\alpha\geq0$, the net effect of the vector field is to \emph{decrease} the rate of expansion of the universe $H$, as compared to one without the vector field, given the same amount of energy density from other matter fields.  Meanwhile, since $\beta_1>0$, from (\ref{beta}) we must have $\delta <0$.  We therefore know from (\ref{gndef}) that the effective gravitational constant is \emph{increased} in the solar system:
\be
  \gc \leq \gs \leq \gn .
\ee
The equalities in this expression only hold if $\alpha$ or $\delta$ vanish. The measured value of Newton's constant in cosmology will thus always be less than expected from Solar-System measurements, so long as $m^2\neq 0$.

Note that since we have not specified the behavior of the matter field except to say that it is a perfect fluid, this remarkable fact is independent of the form of the matter fields. In other words, given any mixture of matter fields, including radiation, dark matter, or any other form of exotic matter (even one with non-constant equation of state parameter), the energy density of the vector field will track it. This is because the energy density (pressure) of the vector field is proportional to the Einstein tensor $00$ ($ij$) term, so the vector field mimics the behaviour of the other matter fields present in the universe leading to a rescaling of the gravitational constant. This rescaling was first mentioned in \cite{Mattingly:2001yd}. 

\section{The Big Bang Nucleosynthesis Speed Limit}

The different rescalings of Newton's constant in the Solar System versus the universe as a whole offer a potential window for observational constraints on the parameters of our vector field.  Newton's constant enters cosmological observations in different ways, including measurements of the expansion rate in the present universe, the formation of late time large-scale structure (e.g \cite{Clifton:2004st}) and the properties of perturbations in the Cosmic Microwave Background.  However, the most straightforward test comes from the predictions of primordial abundances from Big Bang nucleosynthesis (BBN).  Changes to the effective value of Newton's constant are equivalent to the presence of additional (or lack of some) radiation components, leading to a change in the rate of expansion (see {\it e.g.} \cite{Accetta:1990au,Campbell:1995bf,Kolb:1986sj,Uzan:2002vq}).

Decreasing the rate of cosmic expansion during BBN results in weak interactions freezing-out later, leading to a lower freeze-out temperature.  This yields a decrease in the production of primordial $^4\textrm{He}$, and subsequently a lower ${}^4\textrm{He}$ to hydrogen H mass ratio $Y_p$. Current observational bounds on the primordial helium abundance are given by \cite{Olive:1995fe,Izotov:1998mj,O'Meara:2000dh} to be
\be
|\triangle Y_p| < 0.01 ,
\label{ylimit}
\ee 
where the deviation is measured from the value calculated via standard BBN with three massless neutrino species.  Changing the number of effective relativistic degrees of freedom ({\it e.g.} by giving sufficient mass to one or more neutrino species) affects the expansion rate of the universe via the Friedmann equation.  It is therefore convenient to quote limits in terms of a ``cosmic speed-up factor'' $\zeta = H/\bar{H}$, where $\bar{H}$ is the expected (standard) value.  The speed-up factor is related to the helium abundance via \cite{Chen:2000xx}
\begin{equation}
\triangle Y_p = 0.08(\zeta^2-1).
\end{equation}
Our theory predicts a speed-up factor
\be
  \zeta = \sqrt{\gc\over \gn} = \sqrt{{1+8\pi \gs \delta\over
  1 + 8\pi \gs\beta}} , 
\ee
which implies a bound
\begin{equation}
7(2\beta_1+3\beta_2+ \beta_3)m^2 < \frac{1}{8\pi \gn }. \label{BBNCon}
\end{equation}
Notice that the conditions (\ref{conditions}) on the theory parameters $\beta_i$ imply that $(2\beta_1+3\beta_2+\beta_3)>0$, so the left hand side of inequality (\ref{BBNCon}) is always greater than zero (or exactly zero, if $m=0$).

This is our observational bound on the properties of the Lorentz-violating vector field.  Roughly speaking, if the dimensionless parameters $\beta_i$ are of order unity, the norm $m$ of the vector field must be less than $10^{18}$~GeV, an order of magnitude below the Planck scale.  In particular, Planck-scale vector fields are ruled out. 

\section{Conclusions}

We have found solutions to Einstein's equation in the presence of other matter fields for a class of Lorentz-violating, fixed-norm vector field theories, and find that they act to rescale the value of Newton's constant. By comparing these rescalings in the Newtonian regime to those in cosmology, we find an observable deviation from ordinary general relativity.  Following this, we use the predictions of BBN to place constraints on the value of the norm of this vector field. 

Further constraints on this theory, in addition to those already cited in this paper on the $\beta_i$'s, can be derived when we consider its perturbations. In particular, the presence of this vector field during inflation will modify the primordial spectrum of perturbations leading to observable features on both the temperature and polarization anisotropy spectra of the Cosmic Microwave Background. This investigation is the subject of a companion publication \cite{Lim:2004js}.

\section{Acknowledgments}

We would like to thank Christian Armendariz-Picon, Jim Chisholm, Christopher Eling, Wayne Hu, Ted Jacobson, Alan Kostelecky, David Mattingly, Eduardo Rozo, Charles Shapiro, and Susan Tolwinski for useful discussions.  This work was supported in part by the National Science Foundation (Kavli Institute for Cosmological Physics), the Department of Energy, and the David and Lucile Packard Foundation.

\bibliography{slowdown}

\end{document}